\documentclass[twocolumn,showpacs,preprintnumbers,amsmath,amssymb,superscriptaddress,nofootinbib]{revtex4-1}
\usepackage{amsmath,amssymb,bm,graphicx,epsfig,psfrag,ulem,natbib}
\usepackage[usenames, dvipsnames]{color}

\usepackage[abc]{overpic}

\newcommand{\vv}{{\bf {v}}}
\newcommand{\rr}{{\bf {r}}}
\newcommand{\pp}{{\bf {p}}}

\newcommand{\sss}{{\bf {s}}}

\newcommand{\bnabla}{{\mbox{\boldmath $\nabla$}}}

\begin{document}

\title{Visualizing Pure Quantum Turbulence in Superfluid $^{3}$He: Andreev Reflection\\ and its Spectral Properties}

%USE WITH REVTEX4

\author{A.\,W.~Baggaley}
\affiliation{Joint Quantum Centre Durham-Newcastle, and School of Mathematics and Statistics, Newcastle University, Newcastle upon Tyne, NE1 7RU, UK}

\author{V.~Tsepelin}
\email[Electronic Address: ]{v.tsepelin@lancaster.ac.uk}
\affiliation{Department of Physics, Lancaster University, Lancaster, LA1 4YB, UK}

\author{C.\,F.~Barenghi}
\affiliation{Joint Quantum Centre Durham-Newcastle, and School of Mathematics and Statistics, Newcastle University, Newcastle upon Tyne, NE1 7RU, UK}
%\affiliation{Joint Quantum Centre Durham-Newcastle}

\author{S.\,N.~Fisher}
\thanks{Deceased 4 January 2015}
\affiliation{Department of Physics, Lancaster University, Lancaster, LA1 4YB, UK}

\author{G.\,R.~Pickett}
\affiliation{Department of Physics, Lancaster University, Lancaster, LA1 4YB, UK}

\author{Y.\,A.~Sergeev}
\affiliation{Joint Quantum Centre Durham-Newcastle, and School of Mechanical and Systems Engineering, Newcastle University, Newcastle upon Tyne, NE1 7RU, UK}
%\affiliation{Joint Quantum Centre Durham-Newcastle}

\author{N.~Suramlishvili}
\affiliation{Department of Mathematics, University of Bristol, Bristol, BS8 1TW, UK}

\date {\today}

%%%%%%%%%%%%%%%%%%%%%%%%%%%%%%%%%%%%%%%%%%%%%%%%%%%%%%%%%%%%%%%%%%%%%%%%%%%%%%%%%%%%%%%%%%%%

\begin{abstract}

Superfluid $^3$He-B in the zero-temperature limit offers a unique means of studying quantum turbulence by the Andreev reflection of quasiparticle excitations by the vortex flow fields. We validate the experimental visualization of turbulence in $^3$He-B by showing the relation between the vortex-line density and the Andreev reflectance of the vortex tangle in the first simulations of the Andreev reflectance by a realistic 3D vortex tangle, and comparing the results with the first experimental measurements able to probe quantum turbulence on length scales smaller than the inter-vortex separation.
\end{abstract}

\maketitle

Classical turbulence is well known for being simultaneously of universal impact while analytically intractable -- the most important unsolved problem of classical physics as Feynman may have expressed it. One way forward is to start with a simpler system. A pure superfluid in the $T=0$ limit has zero viscosity and can be considered an ideal fluid \cite{Donnelly-book}. While bulk superfluid flow must be irrotational, it {\it can} mimic classical turbulence by supporting singly quantized vortices. At low temperatures, each vortex moves with the local superfluid velocity~\cite{BDV}, comprising the combined velocity fields of all the other vortices~\cite{Donnelly-book,BDV-book}. The system provides a concrete example of the vortex-filament model and the resulting complex flow (a vortex tangle) is quantum turbulence.

Despite the absence of frictional dissipation, quantum turbulence in the $T=0$ limit behaves remarkably similarly to classical turbulence~\cite{Vinen-Niemela} exhibiting a Kolmogorov-like energy spectrum~\cite{NaturePhys2011,PNAS-Kolmo}. Turbulence in superfluid $^3$He-B at microkelvin temperatures provides several advantages over other systems, most importantly that the vortices in this system can be visualized directly by the Andreev reflection of ambient thermal excitations. Such visualization methods have already demonstrated that a vortex tangle forms from the collisions of independent vortex rings~\cite{Bradley2005}, and has begun revealing statistical properties of quantum turbulence~\cite{Bradley2008,FisherPNAS2014}.

Here, we present the first numerical simulations of Andreev reflection by experimentally realistic, three-dimensional vortex tangles in $^3$He-B, and contrast them with the latest experimental measurements of pure quantum turbulence able to probe length scales smaller than the average intervortex distance. This combined numerical/experimental approach allows us to understand the connection between the vortex-tangle line density, $L$, (the total vortex-line length per unit volume) the quantity characterizing the intensity of turbulence, and the reflection coefficient of the thermal excitations, which is used experimentally to visualize the turbulence.

Andreev reflection arises in the $^3$He-B Fermi superfluid as follows. The BCS excitation dispersion curve $E({\bf p})$ has a minimum at the superfluid energy gap, $E_{\rm min}=\Delta$, at the Fermi momenta $p_F$. When an excitation transits from one side of a minimum to the other, its group velocity reverses.  For a superfluid in motion (with velocity $\vv$) the dispersion curve tilts by Galilean transformation to become $E(\pp)+\pp\cdot\vv$~\cite{Fisher_review}. Thus a quasiparticle, moving into a region with superflow parallel to its momentum, experiences a potential barrier. If it has insufficient energy to surmount this barrier then it must be reflected as a quasihole with negligible momentum transfer.  Furthermore, most importantly, the outgoing quasihole almost {\it exactly}\/ retraces the path of the incoming quasiparticle~\cite{BSS1}. (Since any momentum transfer to the superfluid is minimal the excitation motion is rectilinear but the direction can reverse.) Similarly, quasiholes moving into a region of approaching flow will be Andreev-reflected as quasiparticles.  Andreev reflection therefore offers an ideal passive probe for observing vortices at very low temperatures and can provide detailed information about the turbulent behavior.

To set the scene for the simulation, we take a test volume and inject a sequence of vortex rings into it. The rings collide, the cores intersect and recombine, gradually building up an approximately homogeneous tangle.  We then illuminate the tangle with a beam of excitations and calculate the reflection probability. Simulating the simultaneous evolution of the vortex configuration and of the thermal excitations is complicated and numerically expensive. Luckily, we can make several simplifications.  Since the timescale of the quasiparticle motion is much shorter than that of the vortex line motion~\cite{BSS4}, we first obtain from Eqs.~(\ref{eq:biot_savart}) shown below the vortex configuration and associated flow fields, $\vv(\rr,\,t)$ at time $t$, and then analyze the propagation of excitations through this ``frozen'' flow field. Furthermore, the excitation trajectories can be taken as ballistic, since the mean free paths at 175\,$\mu$K greatly exceed any experimental dimension, and, finally, we assume the excitations to be point particles since the coherence length, governing their spatial extent,  $\sim$60\,nm, is tiny compared with the vortex scattering radius, $\sim$20\,$\mu$m.

The superflow field $\vv(\rr,\,t)$ and the dynamics of the vortex tangle are determined by the coupled equations
\begin{equation}
\vv(\rr,\,t)=-\frac{\kappa}{4\pi}\oint_{\cal L}\frac{\rr-\sss}{\vert\rr-\sss\vert^3}\times d\sss\,, \quad \frac{d\sss}{dt}=\vv(\sss,\,t)\,,
\label{eq:biot_savart}
\end{equation}
where the Biot-Savart integral extends over the entire vortex configuration, $\cal L$, $\sss=\sss(t)$ represents a point on the vortex line, and $\kappa=h/2m_3$ (with $m_3$ the mass of a bare $^3$He atom) is the superfluid $^3$He quantum of circulation.

We calculate the superfluid velocity and the time evolution of the vortex tangle using the vortex-filament method with periodic boundary conditions~\cite{BagBar}. To reproduce the experimental situation, see, e.g., Refs.~\cite{Bradley2005,FisherPNAS2014}, we take a cubic box of size $D = 1\,\rm{mm}$ and numerically simulate the evolution of a vortex tangle generated by vortex loop injection for a period of 380\,s. Two rings, radius $R_i= 240$\,$\mu$m, are injected at opposite corners of the numerical domain~\cite{Suppl} at a frequency $f_i= 10$\,Hz. To ensure good isotropy, the loop injection plane is switched at both corners at a further slower rate $f_s= 3.3$\,Hz. The injected vortex loops collide and recombine, rapidly generating a vortex tangle. After an initial transient of $\sim$5\,s, the energy content of the box reaches equilibrium. Losses arise from the numerical spatial resolution limit  ($\approx$6\,$\mu$m), meaning that small scale structures such as high frequency Kelvin waves are lost (but effectively mimicking the loss of kinetic energy from sound radiation at high frequency). The resulting tangle has an equilibrium vortex line density, $\langle L\rangle = 9.7 \times 10^7$\,m$^{-2}$, corresponding to an average intervortex separation of $\ell\approx\langle L\rangle^{-1/2} = 102$\,$\mu$m. The energy spectrum of this tangle is consistent with the $k^{-5/3}$ Kolmogorov scaling for intermediate wave numbers, $k$, and with the $k^{-1}$ scaling for large $k$, see Supplemental Material~\cite{Suppl}.

To analyze the propagation of excitations, an incident quasiparticle flux, in (say) the $x$ direction, is applied normally to one side of the box. The quasiparticle beam is uniformly distributed in the ($y$, $z$) plane and covers the full cross section of the experimental ``cell''. Ignoring angular factors, the incident quasiparticle flux, as a function of position ($y$, $z$), can be written~\cite{FisherPNAS2014}
\begin{equation}
\begin{split}
\langle nv_g\rangle^i_{(y,z)}&=\int^{\infty}_{\Delta}g(E)f(E)v_g(E)dE \\
&= g(p_F) k_B T\exp(-\Delta/k_B T)\,,
\label{eq:Eflux}
\end{split}
\end{equation}
where $g(E)$ is the density of states, and $f(E)$ is the Fermi distribution function, approximated at $T\approx0$ by the Boltzmann distribution $f(E) = \exp(-E/k_BT)$. Since typically quasiparticle group velocities are larger than superflow velocities, the quantity $g(E)v_g(E)$ in integral~(\ref{eq:Eflux}) can be replaced by $g(p_F)$, the density of momentum states at the Fermi energy \cite{FisherPNAS2014}.

In the flow field of the tangle, a quasiparticle (quasihole) moving against (with) a superfluid velocity $\vv$ experiences a force $d{\pp}/dt=-\bnabla(\pp\cdot\vv)$, which pushes it towards the dispersion curve minimum where it becomes a quasihole (quasiparticle) with a reversed group velocity. Consequently, the flux of excitations which can {\it pass}\/ through a tangle is determined by the highest superfluid velocity, $v_x^{\rm max}$, encountered along the excitation's rectilinear trajectory at constant $y$ and $z$, and is thus given by:
\begin{multline}
\langle nv_g\rangle^t_{y,z} = g(p_F)\int^{\infty}_{\Delta + p_Fv_x^{\rm max}} \exp(-E/k_B T) dE \\
= g(p_F)k_BT\exp[{-(\Delta + p_Fv_x^{\rm max})/k_BT}]\,.
\label{eq:Exfluxsimp2}
\end{multline}

The fraction of quasiparticles Andreev reflected by a tangle along the $x$ direction at position ($y$, $z$) is thus
\begin{equation}
f_{y,z} = 1 - \frac{\langle nv_g\rangle^t_{y,z}}{\langle nv_g\rangle^i_{y,z}} = 1 - \exp\left(-\frac{p_Fv_x^{\rm max}}{k_BT}\right)\,.
\label{eq:ExAndsimp}
\end{equation}
The total Andreev reflection $f_x$ is the sum of the Andreev reflections for all positions of the ($y$, $z$) plane. The equivalent calculation is repeated for the thermal quasihole flux and the results combined to give the reflection for a full thermal beam.

\begin{figure}[t]
\begin{center}
\includegraphics[width=8.5cm]{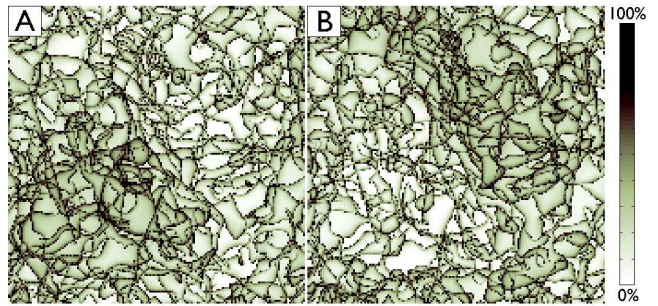}
\caption{(color online). A 2D representation of the reflection coefficient of excitations incident on one side of the calculation cell. Left: quasiparticle reflection. Right: quasihole reflection. The vortex cores are clearly visible as the dark lines. The extended regions of high reflectivity (darker) and low reflectivity (lighter) illustrate the distribution of the large-scale flows in the cell. }
\label{fig:videoshot}
\end{center}
\end{figure}

The simulation~\cite{Simparameters} provides a large volume of information. First, we obtain the magnitude of the Andreev reflection as a 2D contour map across the full cross section of the input excitation beam, see Fig.~\ref{fig:videoshot}, showing very graphically the distribution of
large scale flows across the cell. Unfortunately, experiments do not provide us with similarly detailed information. Therefore, in order to compare theory and experiment, we concentrate instead on the two most significant physical outputs: the average Andreev reflection coefficient, $\langle f_R \rangle$, and, most illuminating, the fluctuations of $f_R$.

\begin{figure}[t]
\begin{center}
\includegraphics[width=7.5cm]{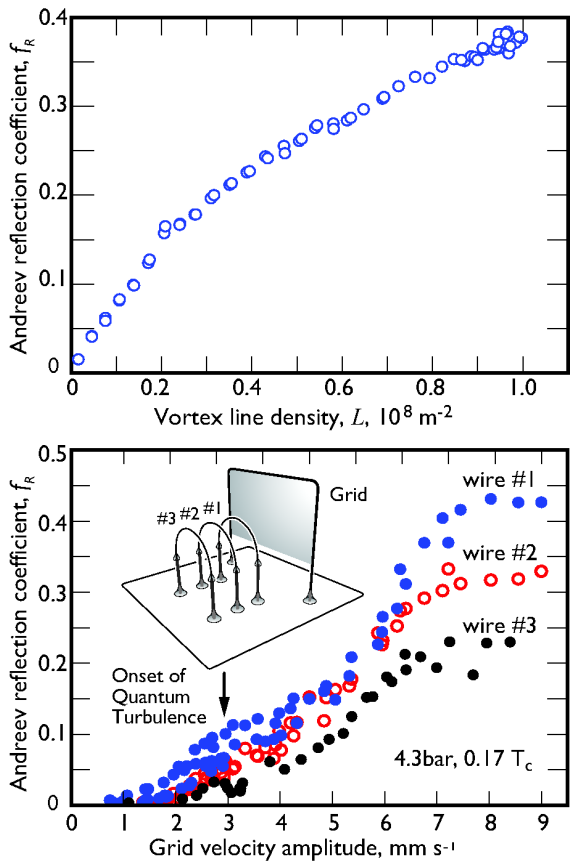}
\caption{(color online). Top: the total reflection coefficient obtained from the numerical simulations, plotted against the line density of the vortex tangle. Bottom: experimental measurements of the fraction of Andreev scattering from vortices generated by a grid plotted versus grid's velocity. Measurements are shown for three vibrating wires at different distances from the grid as shown in the inset, see text.}
\label{fig:combinedfraction}
\end{center}
\end{figure}

The average calculated reflection coefficient is shown in Fig.~\ref{fig:combinedfraction} (top) as a function of the vortex-line density
during the evolution of our tangle. For small line densities, $L\lesssim 2\times10^7$\,m$^{-2}$, the reflection coefficient rises quickly and linearly. At this stage of the tangle's evolution the injected rings are virtually noninteracting. As the simulation progresses, more rings enter the computational domain, start to interact, collide, and form a tangle which absorbs all further injected rings. At higher line densities the rise of Andreev reflection coefficient slows, owing to screening effects. Here we use the term ``screening'' to identify processes which reduce the overall reflectivity of the tangle for a given line density. There may be several mechanisms responsible and further information can be found in the Supplemental Material~\cite{Suppl}.

We compare the simulation with the experiment of Ref.~\cite{Bradley2008}, which has the configuration shown in the inset of Fig.~\ref{fig:combinedfraction}.  The vorticity is produced by the oscillating grid and surrounds the vibrating wire detectors.  The tangle flow fields Andreev reflect ambient thermal excitations arriving from ``infinity'', shielding the wires and reducing the damping. The reduction in damping on each wire (placed at 1.47, 2.37, and 3.49\,mm from the grid) provides the measure of the Andreev reflection by the tangle, see Ref.~\cite{FisherPNAS2014}. The lower part of the figure shows the fractional reflection of quasiparticles incident on each wire.

The numerical and experimental data plots in Fig.~\ref{fig:combinedfraction} have similar shapes. While the vortex line density $L$ cannot be obtained directly from the measurements, we expect the local line density of the quantum turbulence to increase steadily with increasing grid velocity. However, the onset of turbulence is rather different in the simulations and in the experiment. In the simulations, approaching injected vortex pairs are guaranteed to collide and form a tangle, whereas the vibrating grid emits only outward-going vortex rings, with the ring flux increasing steadily with increasing grid velocity. At low grid velocities, rings propagate ballistically with few collisions \cite{Bradley2005,FujiYamaPRB}. At higher velocities, ring collisions increase giving rise to the vortex tangle, which for ther data of Fig.~\ref{fig:combinedfraction} occurs when the  grid velocity exceeds $\sim$ 3\,mm/s. The data at lower velocities correspond to reflection from ballistic vortex rings and can be ignored for the current comparison.

At higher grid velocities or tangle densities, the fraction of excitations Andreev reflected rises at an increasing rate, finally reaching a plateau. The plateau region is prominent in the experiments, and probably results from the extra quasiparticle creation produced when the grid reaches velocities approaching a third of the Landau critical velocity~\cite{LambertPhB}. Compared to the simulations, the absolute value of the reflectivity is almost identical for the wire closest to the grid. This excellent agreement is perhaps fortuitous given that in the experiments quasiparticles travel through 1.5--2.5\,mm of turbulence to reach the wire, compared with 1\,mm in the simulation; thus, larger screening might be expected. A better comparison will require high-resolution experiments to separate the variation of tangle density from the effect of the increasing quasiparticle numbers emitted by the grid.

\begin{figure}[t]
\begin{center}
    \includegraphics[height=6.0cm]{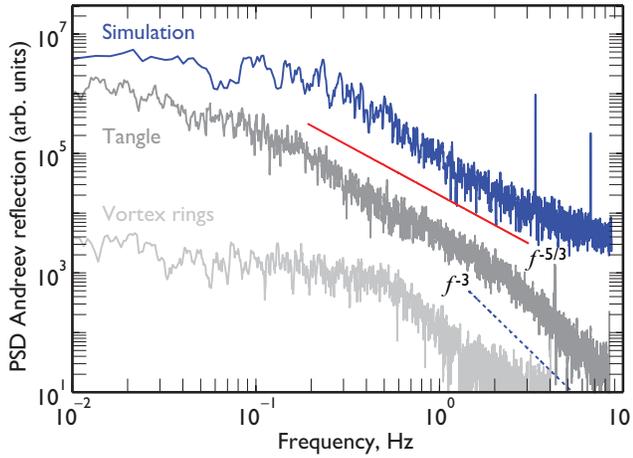}
\caption{(color online). Power spectral density of the Andreev reflection for numerical simulations (top, blue) and experimental observations (middle and bottom, gray) versus frequency. For the details, see text.}
\label{fig:flucsimtanglerings}
\end{center}
\end{figure}

In the simulation, once the tangle has reached the statistically steady state, the vortex line density and the Andreev reflection coefficient fluctuate around their equilibrium, time-averaged values, $\langle L\rangle=9.7\times10^{7}\,{\rm m}^{-2}$ and $\langle f_R\rangle=0.37$, respectively. In order to compare the spectral characteristics of fluctuations of the Andreev reflection $\delta f_R(t)=f_R(t)-\langle f_R\rangle$ and the vortex line density $\delta L(t)=L(t)-\langle L\rangle$ we monitor a steady state of simulated tangle for a period of approximately 380\,s or 7500 snapshots. Taking the Fourier transform $\widehat{\delta f_R}(f)$ of the time signal $\delta f_R(t)$, where $f$ is frequency, we compute the power spectral density $\vert \widehat{\delta f_R}(f) \vert^2$ (PSD) of the Andreev reflection fluctuations. Similarly we compute the PSD $\vert \widehat{\delta L}(f) \vert^2$ of the vortex line density fluctuations. Fig.~\ref{fig:flucsimtanglerings}
shows the Andreev reflection PSD for the simulation (top, blue) and for the experiments (middle and bottom, gray). The experimental data are shown for a fully developed tangle (dark gray) and ballistic vortex rings (light gray) relative to the grid velocities of 6.3\,mm\,s$^{-1}$ and 1.9\,mm\,s$^{-1}$, respectively. (The prominent peaks in the numerical data are artifacts of the discrete vortex-ring injection process.)

The power spectrum of $\delta f_R(t)$ of the simulation and of the experimental data for the developed tangle (reported here and in Ref.~\cite{Bradley2008}) are in excellent agreement showing the same $f^{-5/3}$ scaling behavior at intermediate frequencies. At high frequencies the experimental data develop a much steeper scaling ($\approx\,f^{-3}$), not seen in the numerical spectrum, probably owing to the finite numerical resolution. However, this frequency dependence {\it is} observed in the experiment where only microscopic vortex rings are propagating through the active region and there are no large-scale flows or structures. Thus, we can argue that the $f^{-3}$ scaling for the vortex tangles corresponds to Andreev reflection from superflows on length scales smaller than the intervortex distance.

At a grid velocity of 6.3\,mm\,s$^{-1}$, the tangle propagates at a mean velocity of 0.6--0.8\,mm\,s$^{-1}$\cite{FisherPNAS2014}, Using Taylor's frozen hypothesis, we find that the crossover between the two scaling laws corresponds to a length scale of $\sim$200--300\,$\mu$m, in a good agreement with the intervortex distance obtained from the inferred line density.

\begin{figure}[t]
\begin{center}
    \includegraphics[height=6.5cm]{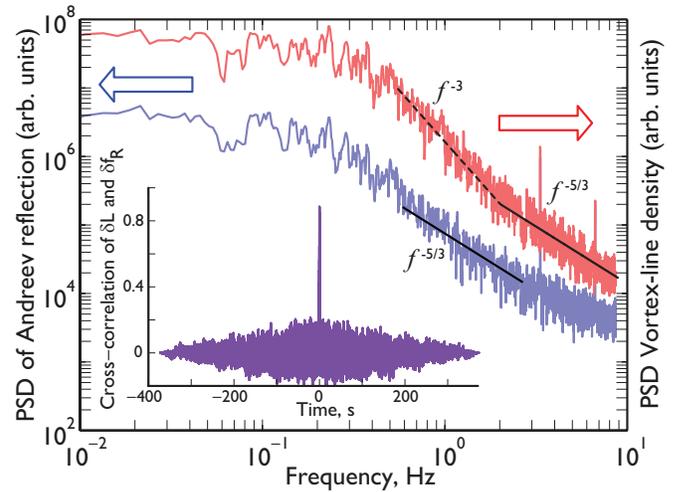}
\caption{(color online). Power spectral densities of simulated Andreev reflection (bottom, blue) and of simulated vortex line density (top, red) versus frequency.  Note that the accuracy of the initially calculated line density PSD extends to higher frequencies than the Andreev-reflection PSD derived from it. Inset: cross-correlation between the Andreev reflection coefficient and vortex line density. The central peak reaches a value of 0.9.}
\label{fig:anddensfluc}
\end{center}
\end{figure}

Finally, we study the relationship between the fluctuations of the vortex line density, $\delta L(t)$, and the fluctuations of the Andreev reflection, $\delta f_R(t)$, by computing the normalised cross-correlations
\begin{equation}
F_{LR}(\tau)=\frac{\langle \delta L(t)\delta f_R(t+\tau) \rangle}
{\sqrt{\langle\delta L^2(0) \rangle}\sqrt{\langle\delta f_R^2(0) \rangle}}\,,
\label{eq:correl}
\end{equation}
where the angle brackets indicate averaging over time, $t$, in the saturated regime, and $\tau$ is the time lag. The insert of Fig.~\ref{fig:anddensfluc} shows that the
cross-correlation between the vortex line density and the Andreev reflection is significant, with $F_{LR}(0)\approx0.9$, clearly demonstrating the link between them, and validating the method of visualization based on Andreev reflection.

Figure~\ref{fig:anddensfluc} contrasts the spectral properties of the vortex line density and of the Andreev reflection from the simulation. At high frequencies, the vortex line spectrum is dominated by the contribution of unpolarized, random vortex lines and exhibits $f^{-5/3}$ scaling~\cite{BagLauBar}. In the intermediate frequency range, this fluctuation spectrum shows $f^{-3}$ scaling and is governed by the large scale flows indicating polarized vortex lines (polarized in the sense of cooperatively correlated), in agreement with recent numerical simulations~\cite{BagLauBar}. If we reasonably assume that this crossover should occur at around the frequency corresponding to the intervortex distance, $\ell$, then using the value of 102$\,\mu$m for $\ell$ calculated above for the equilibrium tangle, the crossover from $f^{-3}$ to the $f^{-5/3}$ behavior should occur at frequency $f_{\ell} \approx v/\ell=\kappa/(2 \pi \ell)\approx1$\,Hz. This is in very fair agreement with the frequency of the crossover between the two regimes of $\approx2$\,Hz as seen in Fig.~\ref{fig:anddensfluc}.

We conclude that the Andreev reflectance of a vortex tangle does indeed reveal the nature of quantum turbulence. The $f^{-5/3}$ scaling of the frequency spectrum of the Andreev-retro-reflected signal has been observed earlier in the experiment of Bradley {\it et al.} ~\cite{Bradley2008}. Starting from simple physical considerations about the flow field in the vicinity of a vortex filament, Bradley {\it et al.} argued that fluctuations of the Andreev-reflected signal can be interpreted as fluctuations of the vortex-line density in turbulent $^3$He-B. The combined numerical-experimental results which we present here show that the fluctuations of the vortex line density and of the Andreev reflection are indeed correlated. However, their spectral densities behave differently. For large scale flows, the vortex line density scales as $f^{-3}$, while the Andreev reflection scales as $f^{-5/3}$. Interestingly, and perhaps coincidentally, the scaling is reversed for an unpolarized tangle. The spectral densities of the Andreev refection fluctuations and of the vortex line density scale as $f^{-3}$ and $f^{-5/3}$, respectively. Thus it is very clear that the Andreev reflection technique has great potential for elucidating the properties of pure quantum turbulence.

This work was supported by the Leverhulme Trust grants F/00 125/AH and F/00 125/AD, and the UK EPSRC grant EP/I028285/1 and the EU FP7 network MICROKELVIN (FP7-228464). Underlying data can be found from Lancaster University research portal at http://dx.doi.org/10.17635/lancaster/researchdata/12.

%\section*{References}


\begin{thebibliography}{99}
\bibitem{Simparameters}
For direct comparison with experiment, we set the temperature to $T=0.15T_c$ and use the appropriate low-pressure $^3$He-B parameters: Fermi momentum $p_F = 8.28 \times 10^{-25}$~kg\,m\,s$^{-1}$,
transition temperature $T_c = 1$\,mK, and energy gap $\Delta=1.76k_BT_c$.
\end{thebibliography}
\end{document}

\title{Supplemental Material\\Visualizing Pure Quantum Turbulence in Superfluid $^{3}$He:\\ Andreev Reflection and its Spectral Properties}

%USE WITH REVTEX4

\author{A.\,W.~Baggaley}
\affiliation{School of Mathematics and Statistics, Newcastle University, Newcastle upon Tyne, NE1 7RU, UK}

\author{V.~Tsepelin}\affiliation{Department of Physics, Lancaster University, Lancaster, LA1 4YB, UK}

\author{C.\,F.~Barenghi}
\affiliation{School of Mathematics and Statistics, Newcastle University, Newcastle upon Tyne, NE1 7RU, UK}\affiliation{Joint Quantum Centre Durham-Newcastle}

\author{S.\,N.~Fisher}\affiliation{Department of Physics, Lancaster University, Lancaster, LA1 4YB, UK}

\author{G.\,R.~Pickett}\affiliation{Department of Physics, Lancaster University, Lancaster, LA1 4YB, UK}

\author{Y.\,A.~Sergeev}
\affiliation{School of Mechanical and Systems Engineering, Newcastle University, Newcastle upon Tyne, NE1 7RU, UK}\affiliation{Joint Quantum Centre Durham-Newcastle}

\author{N.~Suramlishvili}\affiliation{Department of Mathematics, University of Bristol, Bristol, BS8 1TW, UK}

\maketitle

%\centerline{\bf Supplementary material}
%\bigskip

%%%%%%%%%%%%%%%%%%%%%%%%%%%%%%%%%%%%%%%%%%%%%%%%%%%%%%%%%%%%
%%  FIG 1
\begin{figure}[ht]
{\bf (1) SNAPHOT OF THE TANGLE AT THE MOMENT OF RING INJECTION}
\begin{center}
\includegraphics[width=7.5cm]{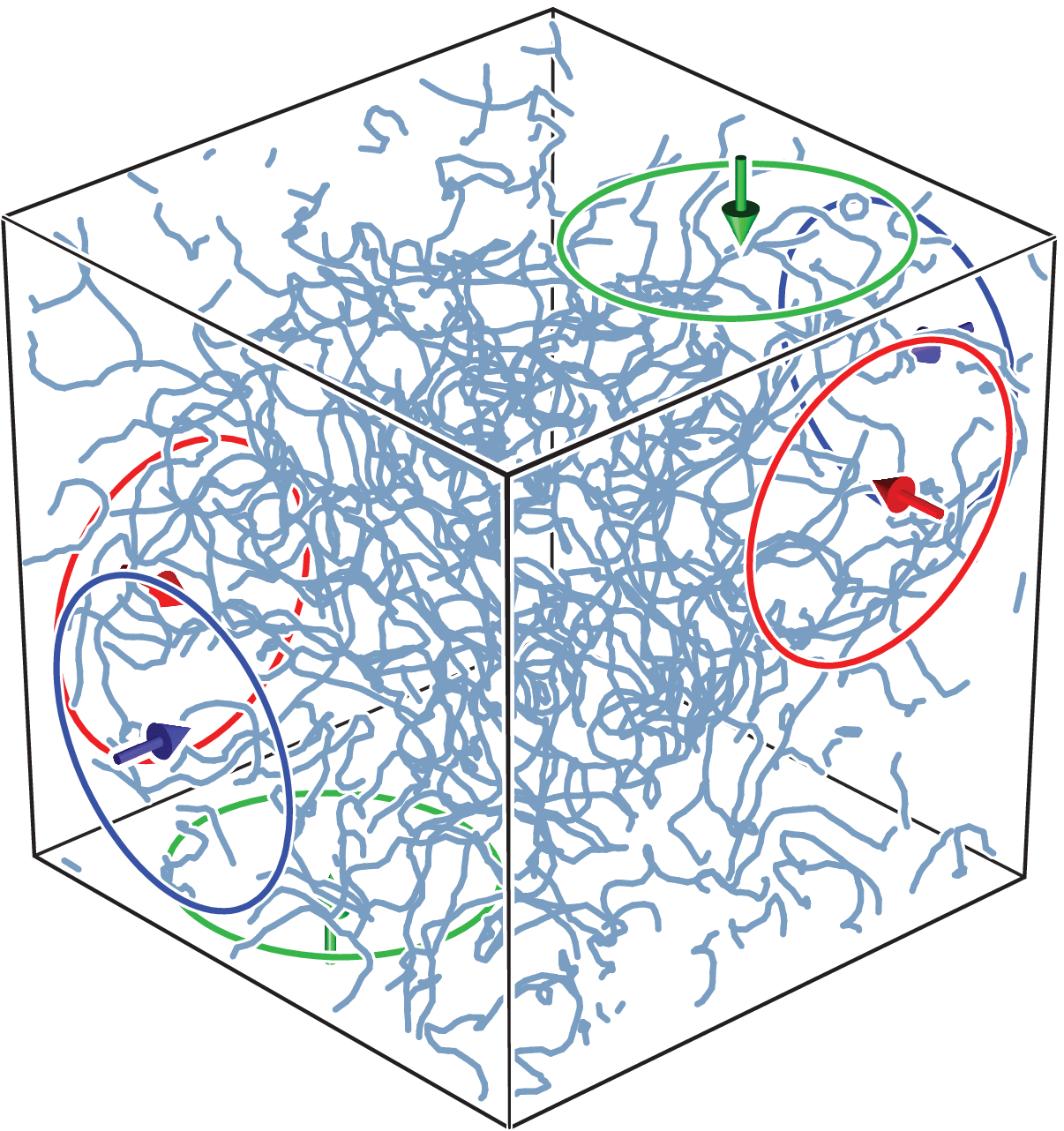}
\caption{(Color online) A snapshot of the tangle at the moment of ring injection from the opposite corners
of the computational domain (note the vortex loops entering the computational box).}
\label{fig:tangle}
\end{center}
\end{figure}
%%%%%%%%%%%%%%%%%%%%%%%%%%%%%%%%%%%%%%%%%%%%%%%%%%%%%%%%%%%%

\begin{figure}[ht]
{\bf (2) ENERGY SPECTRUM OF THE VORTEX TANGLE}
\begin{center}
\includegraphics[width=7.5cm]{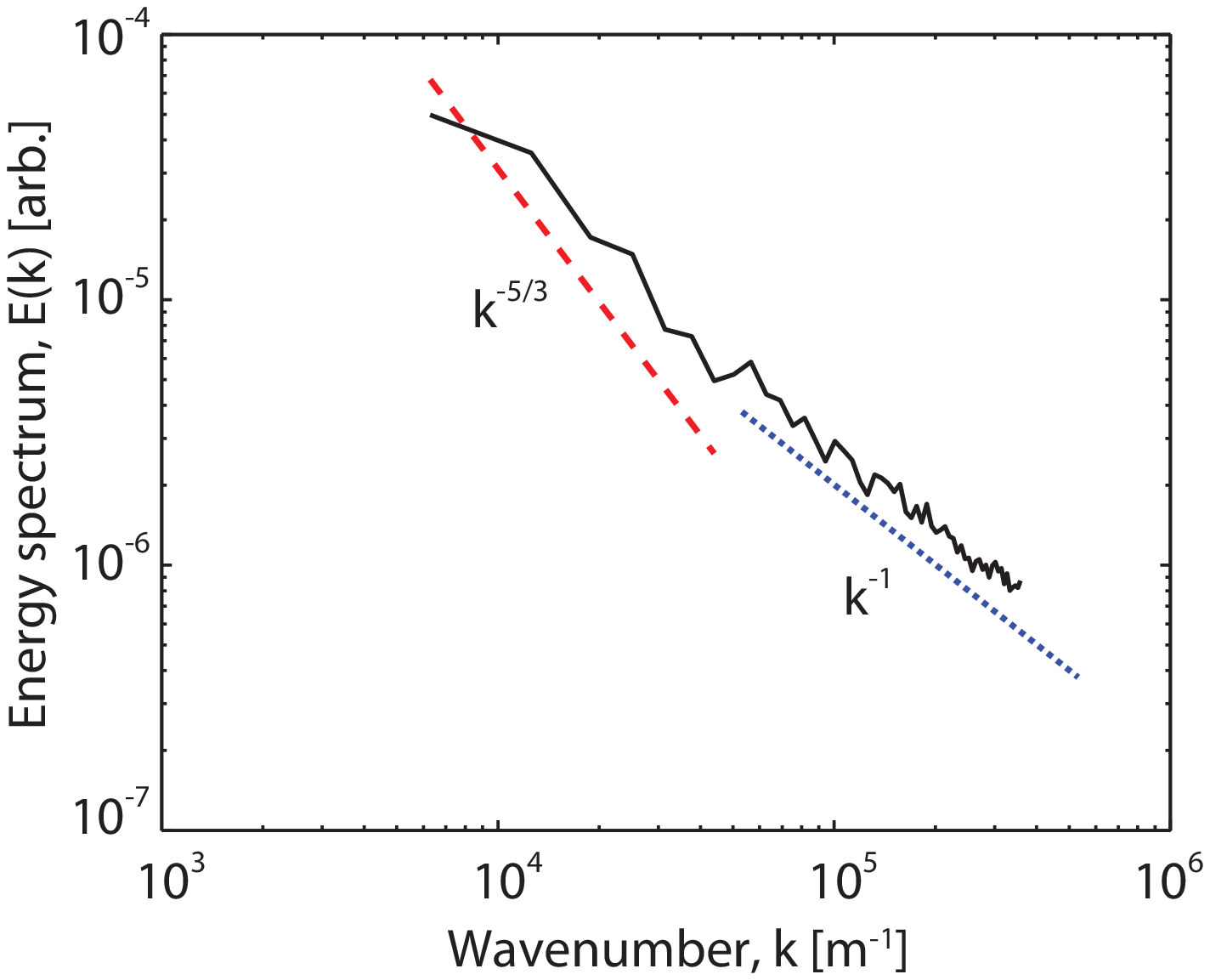}
\caption{(Color online) Energy spectrum of the vortex tangle of the equilibrium line density $\langle L\rangle=9.7\times10^7\,{\rm m}^{-2}$. The slope of the spectrum is consistent with the $k^{-5/3}$ Kolmogorov scaling (red dashed line) for $k<k_{\ell}=2\pi/\ell\approx6\times10^4\,{\rm m}^{-1}$, and with the $k^{-1}$ scaling (blue dotted line) for larger $k$. Here $k$, ${\rm m}^{-1}$ is the wavenumber.}
\label{fig:spectra}
\end{center}
\end{figure}

%%%%%%%%%%%%%%%%%%%%%%%%%%%%%%%%%%%%%%%%%%%%%%%%%%%%%%%%%%%%

\centerline{\bf (3) SCREENING MECHANISMS IN ANDREEV}
\centerline{\bf REFLECTION FROM VORTEX}
\centerline{\bf CONFIGURATIONS AND TANGLES}
\medskip

The analysis of Andreev reflection from an isolated vortex can be found in Ref.~\cite{BSS1}. This has been developed further in Ref.~\cite{SBFTS} to calculate the corresponding scattering length.

The Andreev reflection from a single, rectilinear vortex line is determined by the $1/r$ behavior of the superflow field, where $r$ is the distance from the vortex core. For dilute vortex configurations and tangles the total Andreev reflection is just a sum of reflections from individual vortices. However, as the line density increases this is no longer the case due to screening effects. Here by screening we mean processes which reduce the total Andreev reflectivity of the tangle.

There are two main mechanisms of screening. The first, which we called `partial screening' in our earlier work~\cite{BSS2,BSS3,BSS4}, is due to a modification of the $1/r$ velocity field of the vortex caused, in particular, by neighbouring vortices in its close vicinity. For example, the flow field at some distance from two close, (nearly) antiparallel vortex lines is that of a vortex dipole, $v\sim1/r^2$; this, faster decay with $r$ of the superflow velocity significantly reduces the total reflectivity of the vortex pair. The $1/r$ velocity field may also be modified by the effects of large curvature of the vortex line itself; thus, a reduction of the Andreev reflection from small vortex rings was reported in Ref.~\cite{BSS4}. It can also be expected that the reflectivity of the vortex line may be reduced by the large-amplitude Kelvin waves.

The other mechanism is that of `fractional' screening. This mechanism is of particular importance for large/dense tangles: vortices at the front (with respect to the incident beam of excitations) of the dense or/and large tangle will obscure those at the rear. A simple model~\cite{Bradley2006} of this screening mechanism was developed to yield a rough estimate for the vortex line density from the experimentally measured Andreev reflection (this mechanism was also called `geometric' screening in Ref.~\cite{FisherPNAS2014}). This screening mechanism will also be important in the case where, in dense tangles, vortex bundles~\cite{BagLauBar} may form producing high reflectivity regions surrounded by those whose reflectivity is much lower. This will generate a significant `geometric' screening since few excitations will reach vortices at the rear of a large vortex bundle, and adding a vortex to a region where there is already an intense reflection will not have a significant effect on the total reflectivity.